\documentclass[aps,prl,twocolumn,superscriptaddress]{revtex4}
\usepackage{amsmath,amssymb}
\usepackage{graphics,graphicx,color}
\usepackage{dcolumn,bm,ulem}

\newcommand{\iisc}
{\affiliation{Centre for Condensed Matter Theory, Department of Physics, Indian Institute of Science, Bangalore 560012, India}}
\newcommand{\sheffield}
{\affiliation{Department of Physics and Astronomy, Sheffield University, Hounsfield Road, Sheffield S3 7RH,United Kingdom.}}

\newcommand{\gottingen}
{\affiliation{Institut f\"ur Theoretische Physik, Georg-August Universit\"at G\"ottingen, 37077 G\"ottingen, Germany}}

\begin{document}
\title
{Complex Dynamics of a Model of Sheared Nematogenic Fluids}

\author{Rituparno Mandal}%
\email[Email: ]{rituparno@iisc.ac.in}
\gottingen \iisc

\author{Buddhapriya Chakrabarti}%
\email[Email: ]{b.chakrabarti@sheffield.ac.uk}
\sheffield

\author{Debarshini Chakraborti}%
\email[Email: ]{cdebarshini@ccamp.res.in}

\author{Chandan Dasgupta}%
\email[Email: ]{cdgupta@iisc.ac.in}
\iisc

\begin{abstract}
Nonlinearities in constitutive equations of extended objects in shear flow lead to novel phenomena, {\it e.g.} ``rheochaos'' in solutions of wormlike micelles and ``elastic turbulence'' in polymer solutions. Since both phenomena involve anisotropic objects, their contributions to the deviatoric stress are likely to be similar. However, these two fields have evolved rather independently and an attempt at connecting these fields is still lacking. We show that a minimal model in which the anisotropic nature of the constituting objects is taken into account by a nematic alignment tensor field reproduces several statistical features found in rheochaos and elastic turbulence. We numerically analyse the full non-linear hydrodynamic equations of a sheared nematic fluid under shear stress and strain rate controlled situations, incorporating spatial heterogeneity only in the gradient direction. For a certain range of imposed stress and strain rates, this extended dynamical system shows signatures of \textit{spatiotemporal chaos} and \textit{transient shear banding}. In the chaotic regime the power spectra of the order parameter stress, velocity fluctuations and the total injected power show power law behavior and the total injected power shows a non-gaussian, skewed probability distribution. These dynamical features bear resemblance to \textit{elastic turbulence} phenomena observed in polymer solutions. The scaling behavior is independent of the choice of shear rate/stress controlled method.
\end{abstract}
\maketitle

\section{introduction}
Sheared complex fluids often exhibit unusual features in their dynamics \cite{rev1,rev2,rev3}. For example, sheared solutions of wormlike micelles show complex dynamics~\cite{Becu:04,Lopez:04,Herle:07,Fardin:09,Nghe:10} including spatiotemporal chaos \cite{Ranjini:00,Sood:06,Sood:08} at very low Reynolds numbers. This phenomenon is referred to as ``rheochaos''. Sheared polymeric liquids show irregular flow behavior with fluid motion excited over a large range of spatial and temporal scales~\cite{Steinberg1,Steinberg2,Steinberg3,Steinberg4}. This behavior, akin to turbulent phenomena in Newtonian fluids, is termed ``elastic turbulence''. Both these phenomena show similar statistical properties~\cite{Steinberg1,Steinberg2,Sood:11,Fardin:10,Beaumont:13}, \textit{e.g.} power law decay of the power spectral density (PSD) of fluctuating quantities such as shear rate/shear stress, depending on the controlling protocol. In addition, experiments on elastic turbulence show skewness and non-Gaussianity of the probability distribution function (PDF) of total injected power~\cite{Labbe:99,Steinberg3,Steinberg4,Sood:11}. Since these phenomena occur at low Reynolds numbers, the inertial nonlinearities familiar from Navier-Stokes turbulence are not expected to play an important role. It has been suggested~\cite{Rienacker:02,Rienacker2:02,Chakrabarti:04,Das:05} that nonlinearities in the constitutive relations appearing in the hydrodynamic equations of motion of complex fluids are responsible for the occurrence of some of these phenomena.

In this paper, we explore the role of material nonlinearities in the dynamics of driven complex fluids by numerically analyzing the equations of nematic hydrodynamics under (a) shear rate, and (b) shear stress controlled conditions. Previous numerical studies implementing a shear rate controlled protocol with spatial heterogeneity and passive advection~\cite{Chakrabarti:04,Das:05} as well as incorporating velocity feedback~\cite{Chakraborty:10} showed presence of spatiotemporal chaos. We have extended these calculations to the shear stress controlled protocol and studied the behaviour of several fluctuating quantities measured in experiments on elastic turbulence. The main results of the present study can be summarized as follows: \textit{(i)} while the temporal dynamics of the nematic order parameter of a homogeneous system with stress control~\cite{Klapp:10} does not show chaos, the full hydrodynamic model incorporating spatial heterogeneity shows spatiotemporal chaos in this case also, \textit{(ii)} statistical quantities \textit{e.g.} the spatial and temporal power spectra of the order parameter stress show power law scaling with the wave vector and frequency, with identical exponent values for both control protocols, and \textit{(iii)} the statistical properties of the spatiotemporally chaotic phase found in this model bear a strong similarity with those characterizing elastic turbulence in polymer solutions~\cite{Steinberg2} and micellar solutions\cite{Sood:11,Fardin:10}. These results show that several aspects of the ``chaotic'' or ``turbulent'' behaviour observed in driven complex fluids 
can be reproduced in simple models in which the complex behaviour arises from the presence of  material constitutive nonlinearities in the hydrodynamic equations for the fluid. These results also suggest an intriguing connection between rheochaos and elastic turbulence. While our model is too simple to provide a realistic description of experimental results on either rheochaos or elastic turbulence, our results point out that nonlinearities in the constitutive relations can indeed lead to behaviour similar to those observed in experiments.

\section{Model and Simulation Method}

We have studied a phenomenological model, proposed by Hess~\cite{Hess:7576} that considers the hydrodynamic relaxation equation of the alignment tensor $\textbf{Q}$ of a complex nematogenic fluid incorporating spatial heterogeneities~\cite{Chakrabarti:04,Das:05}. The equation of motion obeyed by the nematic alignment tensor $\textbf{Q}$ is
\begin{equation}
\frac{\partial \textbf{Q}}{\partial t} + \textbf{u}.\mbox {\boldmath $\nabla Q$} = \tau^{-1} \textbf{G} +(\alpha_{0} \boldsymbol{\kappa} +\alpha_{1} \boldsymbol{\kappa}.\textbf{Q})_{ST}+\textbf{Q}.\mathbf{\Omega}-\mathbf{\Omega}.\textbf{Q},\label{eq:OP-dynamics}
\end{equation}
where, $\tau$ is a ``bare'' relaxation time, $\alpha_{0}$ and $\alpha_{1}$ are flow alignment parameters related to molecular shapes, the subscript `ST' denotes symmetrization and trace removal of the tensorial quantities and, and  \mbox{\boldmath $\kappa$} $\equiv (1/2)[\mbox{\boldmath $\nabla u+(\nabla u)$}^T ]$ and $\mathbf{\Omega} \equiv(1/2)$ $[\mbox{\boldmath $\nabla u -(\nabla u)$}^T]$ are the shear rate and vorticity tensors respectively.
The imposed flow geometry is plane Couette type with velocity ${\bf{u}}=y \Gamma \hat{\bf{x}}+\delta_{1} \hat{\bf{x}}+\delta_{2} \hat{\bf{z}}$ where $\delta_{1}$ and $\delta_{2}$ are $y$-dependent perturbations in the velocity profile and $\Gamma$ is the shear rate. Therefore the flow is along the $x$ axis, the $z$ axis is the vorticity direction and spatial variations are allowed only in the gradient direction (the $y$ axis). Thus, we effectively have a quasi-one-dimensional hydrodynamic model for the nematic order parameter and velocity fields.

In Eq.(\ref{eq:OP-dynamics}), $\textbf{G}$ is the molecular field conjugate to $\textbf{Q}$ \textit{i.e.} $\textbf{G}=-\frac{\delta F[\textbf{Q}]}{\delta \textbf {Q}}$ where $ F[\textbf{Q}]$ is the Landau-De Gennes free energy functional,
\begin{eqnarray}
F[\textbf{Q}] &=& \int d^{3}x[\frac{A}{2}\textbf{Q}:\textbf{Q} - \sqrt{\frac{2}{3}}B(\textbf{Q}\cdot\textbf{Q}):\textbf{Q} + \frac{C}{4}(\textbf{Q}:\textbf{Q})^{2} \nonumber\\
&+&\frac{\Gamma_1}{2}\mbox{\boldmath $ \nabla Q$}\ \vdots\ \mbox{\boldmath $\nabla Q$ }
+\frac{\Gamma_2}{2}\mbox{\boldmath $\nabla.Q$}.\mbox{\boldmath $\nabla.Q $}],\label{eq:deGennesF}
\end{eqnarray}
with phenomenological constants $A$, $B$, $C$ controlling the free energy difference between the isotropic and nematic phases and $\Gamma_{1}$ and $\Gamma_{2}$ related to the Frank elastic constants. The cubic and quartic terms in the free energy functional lead to nonlinear terms in the equation of motion, Eq.(\ref{eq:OP-dynamics}), via the molecular field $\textbf{G}$.

The total stress ${\boldsymbol{\sigma}}$ in the system is the sum of  bare viscous stress ${\boldsymbol{\sigma}}^{vis}$ and order parameter stress ${\boldsymbol{\sigma}}^{OP}$. The viscous stress is of the form ${\boldsymbol{\sigma}}^{vis}=\mu [({\boldsymbol {\nabla}} \textbf{u}) + ({\boldsymbol{\nabla}} \textbf{u})^{T}]$ where $\mu$ is the viscosity of the nematogenic fluid and the deviatoric order parameter stress ${\boldsymbol{\sigma}}^{OP}$ can be written as
\begin{equation}
{\boldsymbol{\sigma}}^{OP} =  -\alpha_{0} {\bf{G}}-\alpha_{1}({\bf{Q.G}})_{ST}.\label{eq:OPstress}
\end{equation}
We assume that the fluid is incompressible \textit{i.e.} $\boldsymbol \nabla .\bf{u}=0$ and work in the zero Reynolds number or Stokesian limit \textit{i.e.} $\boldsymbol \nabla .{\boldsymbol{\sigma}}=\bf 0$.

When spatial variations are allowed only in the gradient direction, the Stokes condition enforces $\boldsymbol{\sigma}$ to be constant 
in space. In the shear rate controlled case we implement this constraint by imposing
\begin{equation}
\mu \frac{\partial^2 u_i}{\partial y^2}=-\frac{\partial \sigma^{OP}_{yi}}{\partial y},\label{eq:straincontrol}
\end{equation}
where $i=x,z$ for the specific flow geometry considered.
For the shear stress controlled case $\boldsymbol{\sigma}$ is constant 
in space and time and this restriction is incorporated by imposing
\begin{equation}
\mu \frac{\partial u_i}{\partial y}=\sigma^{imp}_{yi}- \sigma^{OP}_{yi},\label{eq:stresscontrol}
\end{equation}
where $\boldsymbol{\sigma^{imp}}$ is the imposed shear stress. The velocity perturbations $\delta_1$ and $\delta_2$ are determined from Eqs. 
\ref{eq:straincontrol} and \ref{eq:stresscontrol}.

Recently Klapp and Hess~\cite{Klapp:10} have implemented a protocol to study stress controlled rheology of nematogenic fluids. In this model the instantaneous shear stress $\sigma_{xy}$ is expressed in terms of a time-dependent shear rate $\Gamma(t)$ and the deviatoric stress contribution from the order parameter tensor (Eq.~\ref{eq:OPstress}). The rate of change of the shear rate $\Gamma(t)$ with respect to time is set to be proportional to the difference between the instantaneous and imposed stresses. This ensures stress control for times $t > \tau_{g}$, where $\tau_{g}$ is a time scale over which the ``control'' takes effect.
The controlling protocol implemented in this paper does not suffer from this limitation and (Eq.~\ref{eq:stresscontrol}) ensures that the shear stress is exactly equal to the imposed value at each time instant.

The order parameter $\textbf{Q}$ and its time evolution (Eq.~\ref{eq:OP-dynamics}) can be expressed in a orthonormal basis with five independent components $a_0, a_1, \ldots a_4$~\cite{Rienacker:02,Chakrabarti:04}. These equations along with Eq.~\ref{eq:straincontrol} and Eq.~\ref{eq:stresscontrol} (depending on the controlling protocol) provide the full hydrodynamic description of a sheared nematic fluid, which we solve numerically.
We have rescaled space by the diffusion length constructed from $\Gamma_1$ and $A_{*}$, time by $\tau/A_{*}$  and $\textbf{Q}$ by $Q_{k}$ where $A_{*}=\frac{2 B^2}{9 C}$ and $Q_{k}$ is the magnitude of $\textbf{Q}$ at the transition temperature.
The equations of motion of the nematic director have several independent parameters: $A$, $\Gamma$, $\Gamma_{1}$, $\Gamma_{2}$, $\alpha_{1}$, $\lambda_{k}$ and $\eta$ where $\lambda_{k}=\sqrt{\frac{2}{3}} \frac{\alpha_{0}}{Q_{k}}$ and $\eta=\mu/(\alpha_{0} \tau Q_{k})$. The results presented here are for $\Gamma_{2}=\Gamma_{1}=1$, $A=0$, $\alpha_1=0$  and $\eta=1$. We have verified that our results are insensitive to small changes in these parameter values. 
We set ${\sigma^{imp}_{yz}=0}$ in the stress controlled case. Thus we have two independent parameters: the imposed shear stress ${\sigma^{imp}_{xy}}$ or the shear rate ($\Gamma$) and the tumbling parameter ($\lambda_{k}$).

The equation for the time evolution of the director field, Eq.~\ref{eq:OP-dynamics} expressed in terms of the components $a_0, a_1 \ldots a_4$, is solved numerically. A symmetrized finite difference scheme is used to compute the spatial derivatives~\cite{Chakraborty:11} while the equations are integrated forward in time using a fourth-order Runge-Kutta scheme with a fixed time-step $\Delta t = 0.001$. The fluid velocity is calculated using Eq.~\ref{eq:straincontrol} or Eq.~\ref{eq:stresscontrol} at each time step and fed back in Eq.~\ref{eq:OP-dynamics} to compute the instantaneous order parameter profile. The fluid velocity for the shear rate controlled case is obtained by expressing Eq.~\ref{eq:straincontrol} as a matrix equation and performing a matrix inversion using a LAPACK subroutine.

Fixed boundary conditions are implemented for the velocity field $\textbf{u}$ and the order parameter field $\textbf{Q}$, with the $\textbf{Q}$ tensor corresponding to the nematic director being along $\hat{z}$ at the walls (\textit{i.e.} at $y=0$ and $y=L$). We have checked that the behavior in the bulk does not depend on the alignment of the nematic director at the boundaries. The boundary conditions for the velocity field are $\delta_{1}=\delta_{2}=0$ at both the boundaries (which ensures steady shear) for the shear rate controlled case while $\delta_{1}=\delta_{2}=0$ only at the static boundary 
for the shear stress controlled situation.
We have studied system sizes varying between $L=100$ to $10000$ with grid size $\Delta x=0.1$. We have verified that smaller values of the grid size or the integration time step do not change the results significantly. 

\section{Results}

\begin{figure}
\includegraphics[width=8.5cm,height=6cm]{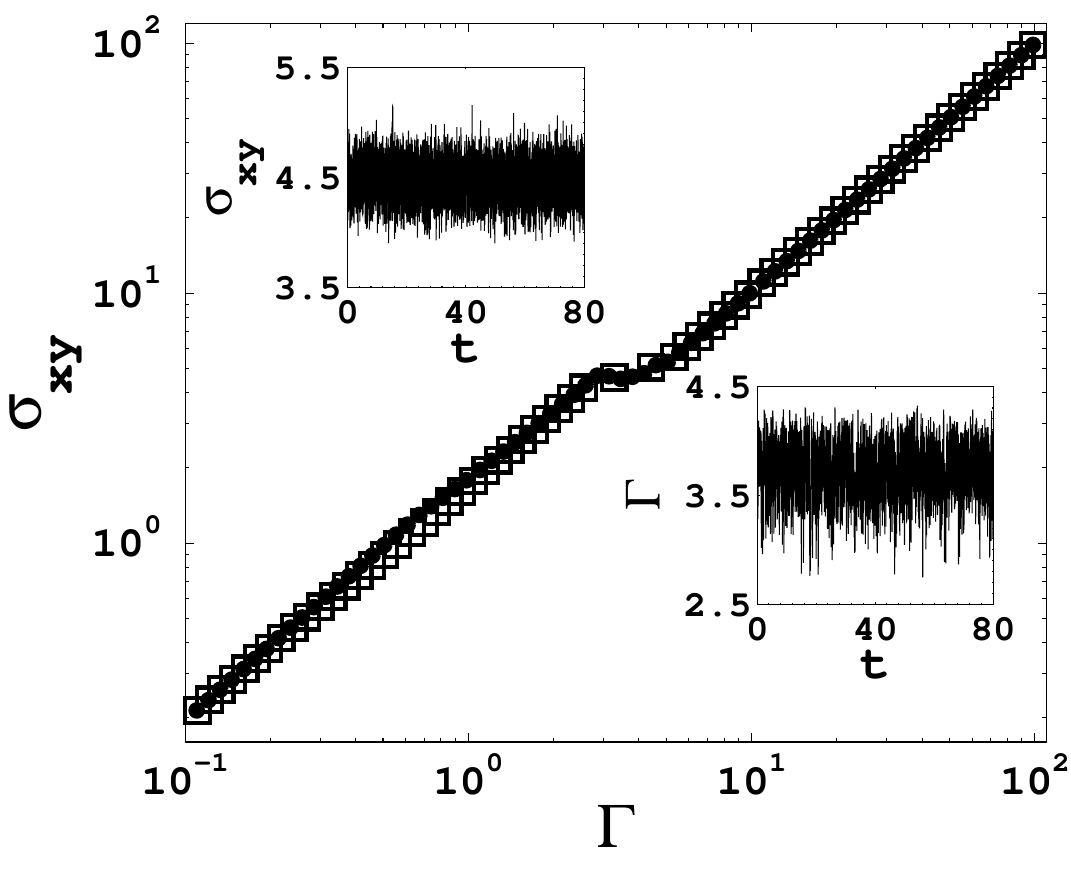}
\caption{Flow curve for the parameter values $\lambda_k=1.12$, $\eta=1.0$, in shear rate (filled circles) and shear stress (open sqares) controlled simulations. Insets show the fluctuation in the shear stress $\sigma_{xy}$ (shear rate $\Gamma$) with respect to time $t$ (in units of $1000$ computational time units), when the shear rate (shear stress) is held constant. }
\label{fig:1}
\end{figure}

\begin{figure}
\includegraphics[width=9cm,height=6.5cm]{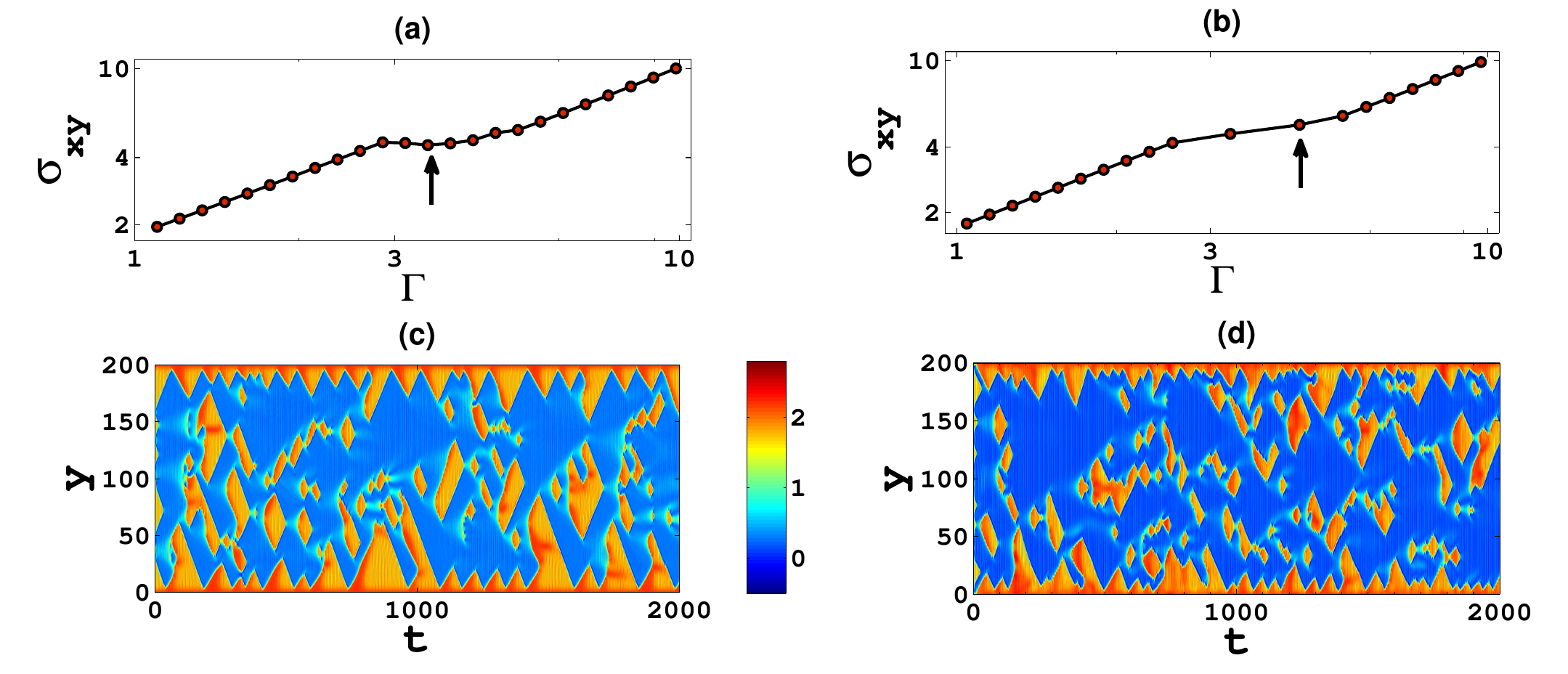}
\caption{Flow curve for (a) shear rate controlled and (b) shear stress controlled protocols for parameters $\lambda_k=1.12$, $\eta=1.0$ and corresponding space time plots for the order parameter stress $\sigma^{OP}_{xy}$ (see text) (panels (c) and (d), respectively) for the unstable part of the flow curve indicated by an arrow. Same color bar has been used for both the plots in panels (c) and (d).}
\label{fig:2}
\end{figure}

\begin{figure}
\includegraphics[width=9cm,height=6cm]{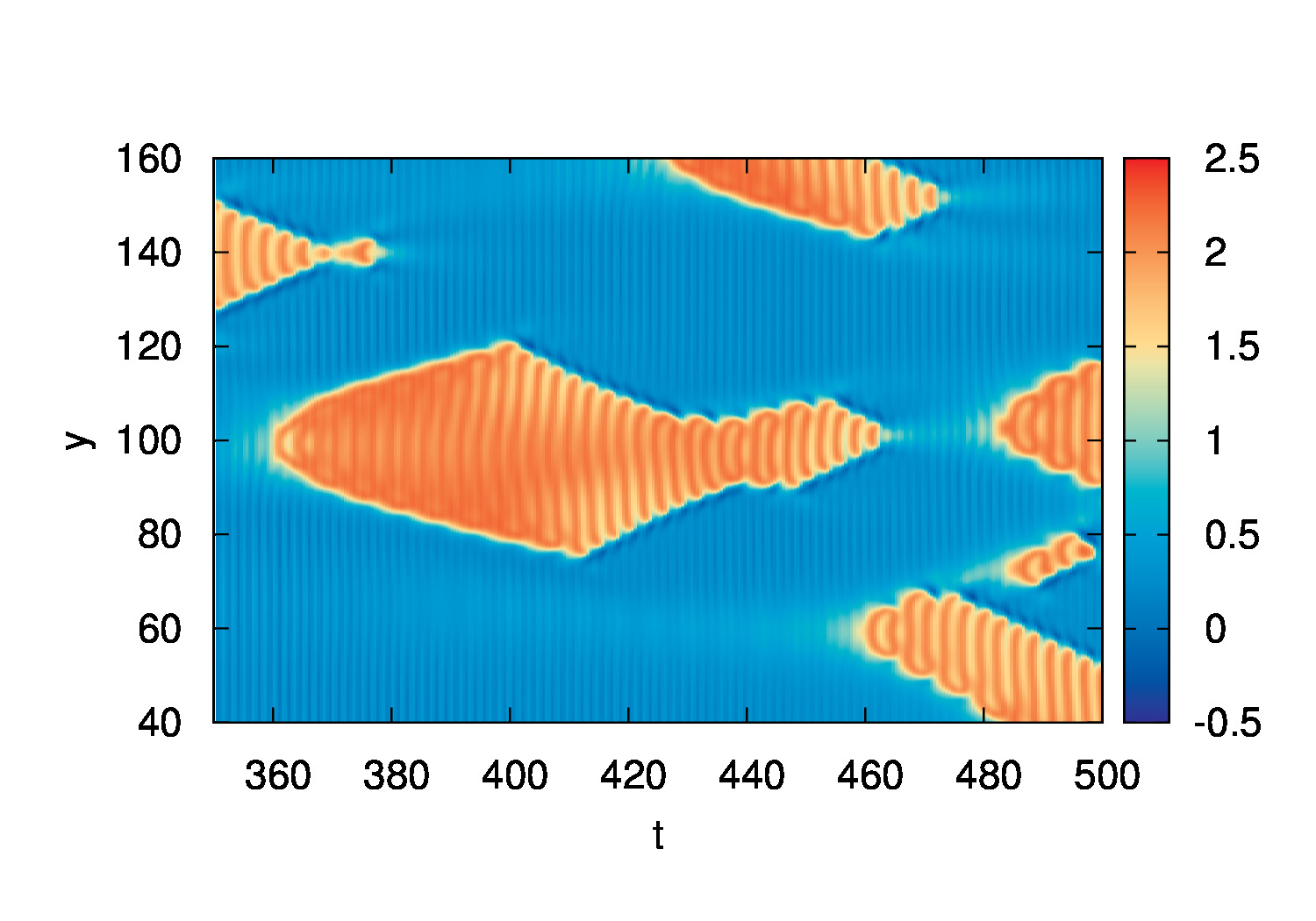}
\caption{Space time plot of order parameter stress component $\sigma_{xy}$ in the spatio temporally chaotic phase for $\lambda_k=1.12$, $\eta=1.0$.}
\label{fig:s1}
\end{figure}

\begin{figure}
\includegraphics[width=9cm,height=6cm]{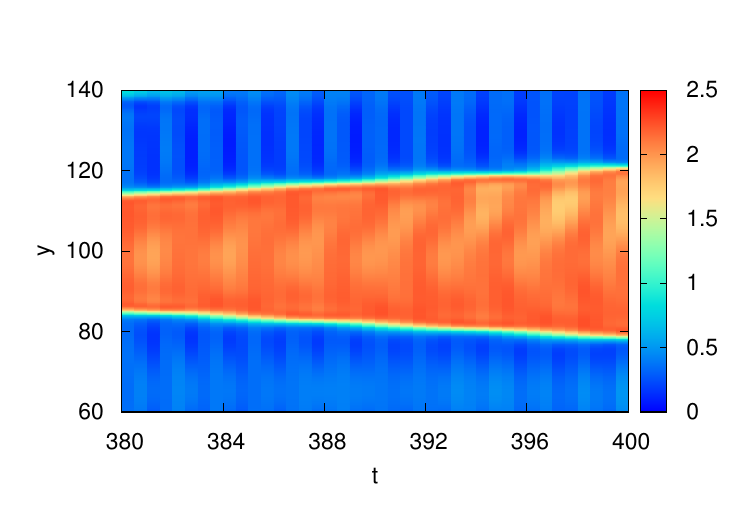}
\caption{Space time plot of order parameter stress component $\sigma_{xy}$ in the spatio temporally chaotic phase for $\lambda_k=1.12$, $\eta=1.0$. Enlarged version of the previous figure.}
\label{fig:s2}
\end{figure}

\begin{figure}
\includegraphics[width=9cm,height=6cm]{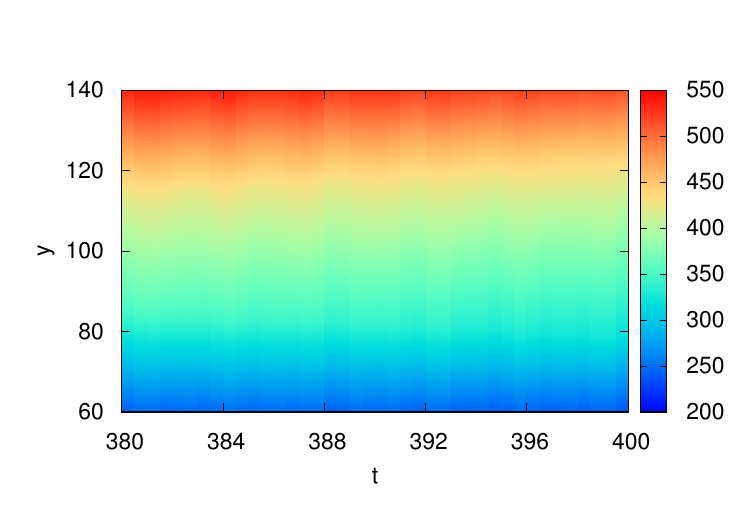}
\caption{Space time plot of velocity $v_{x}$ in the spatio temporally chaotic phase for $\lambda_k=1.12$, $\eta=1.0$.}
\label{fig:s3}
\end{figure}

\begin{figure}
\includegraphics[width=9cm,height=5cm]{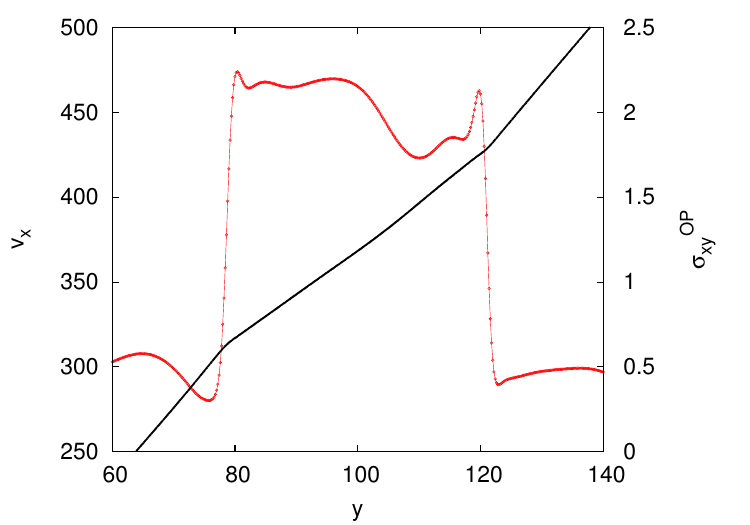}
\caption{Spatial profile of velocity $v_{x}$ (black) and order parameter stress component $\sigma_{xy}$ (red) taken at $t=400$ in the spatio temporally chaotic phase for $\lambda_k=1.12$, $\eta=1.0$.}
\label{fig:s4}
\end{figure}

By keeping $\lambda_{k}$ fixed between $0.9$ to $1.15$ and varying the shear rate $\Gamma$ or the shear stress ${\sigma^{imp}_{xy}}$, we find three distinct steady states or ``phases'' (i) \textit{periodic}, (ii) \textit{spatiotemporally chaotic}, (see Fig.~\ref{fig:2}) and (iii) \textit{aligned}. In the shear rate controlled case, the flow curve (see Fig.~\ref{fig:1}) shows non-monotonic behavior in a small range of $\Gamma$ values, whereas the shear stress controlled flow curve shows discontinuous changes in the shear rate when the shear stress is changed by only a small amount. 
On closer inspection, this region reveals the existence of a spatiotemporally chaotic phase. A detailed investigation of the order parameter stress and velocity profile in the spatiotemporally chaotic phase shows transient shear banding with randomly nucleating domains of low and high stress (see Figs.~\ref{fig:2}(c) and (d) ) that evolve as a function of time. This behavior is shown in detail in
Fig.~\ref{fig:s1} which is a space time plot for the order parameter stress component $\sigma_{xy}$ for the spatiotemporally chaotic phase for $\lambda_k=1.12$, $\eta=1.0$. Fig.~\ref{fig:s2} shows the enlarged version of a section of
Fig.~\ref{fig:s1} and and Fig.~\ref{fig:s3} shows the velocity (along $x$ direction) profile in that specific space time region. Plots of the velocity $v_{x}$ and order parameter stress component $\sigma_{xy}$ as a function of space $y$, along the gradient direction, at time instant $t=400$, shown in Fig.~\ref{fig:s4} show shear banding in both the quantities. However, the space time plots of these quantities clearly show that the banding is not steady in time - rather it is transient in nature.

\begin{figure}
\includegraphics[width=8.5cm,height=6cm]{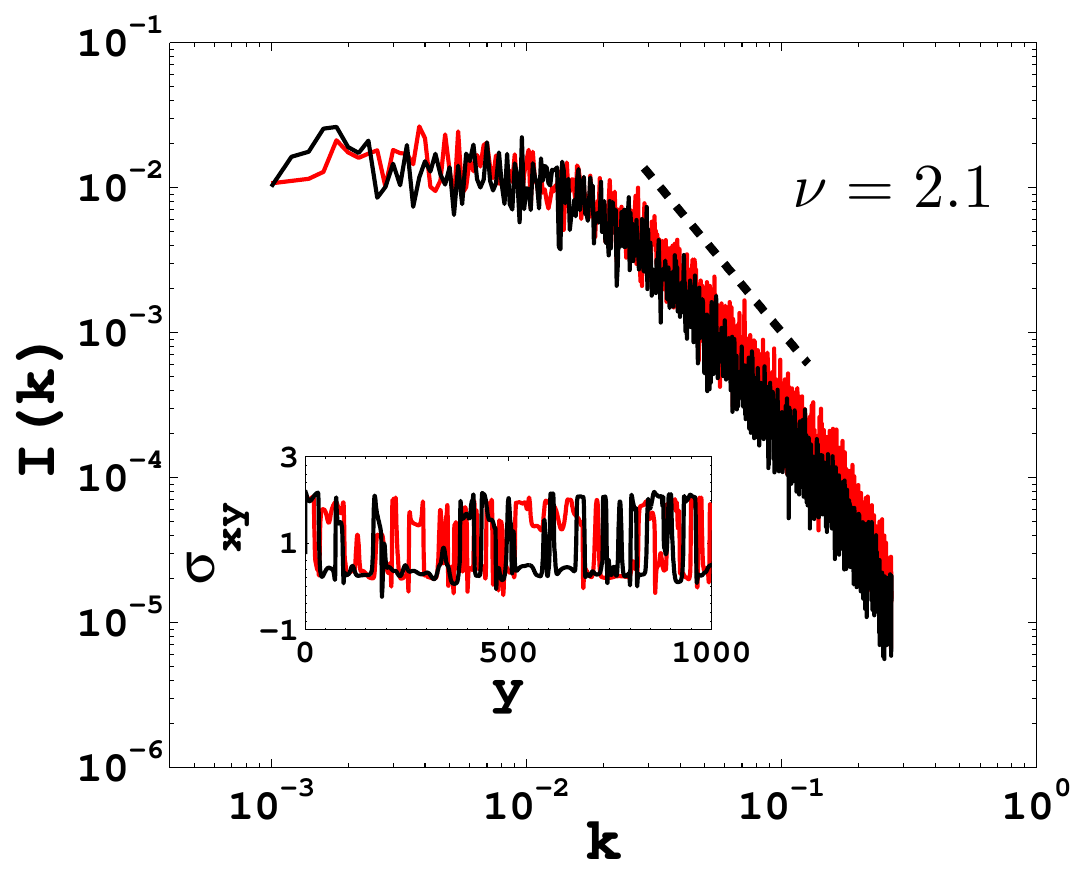}
\caption{Spectrum in the wave vector domain, obtained by taking the Fourier transform of the spatial profile of the order parameter stress, averaged over $10$ 
time instances (separated by a time scale longer than the correlation time) in both shear stress (black) and shear rate (red) controlled simulation for $\lambda_k=1.12$. The dashed line is a power law fit of the form $I(k)\sim k^{-\nu}$ with $\nu=2.1$. The inset shows the spatial variation of the order parameter stress in shear rate (red) and shear stress (black) controlled simulation at an arbitrary time instance in the steady state.}
\label{fig:3}
\end{figure}

\begin{figure}
\includegraphics[width=8.5cm,height=6cm]{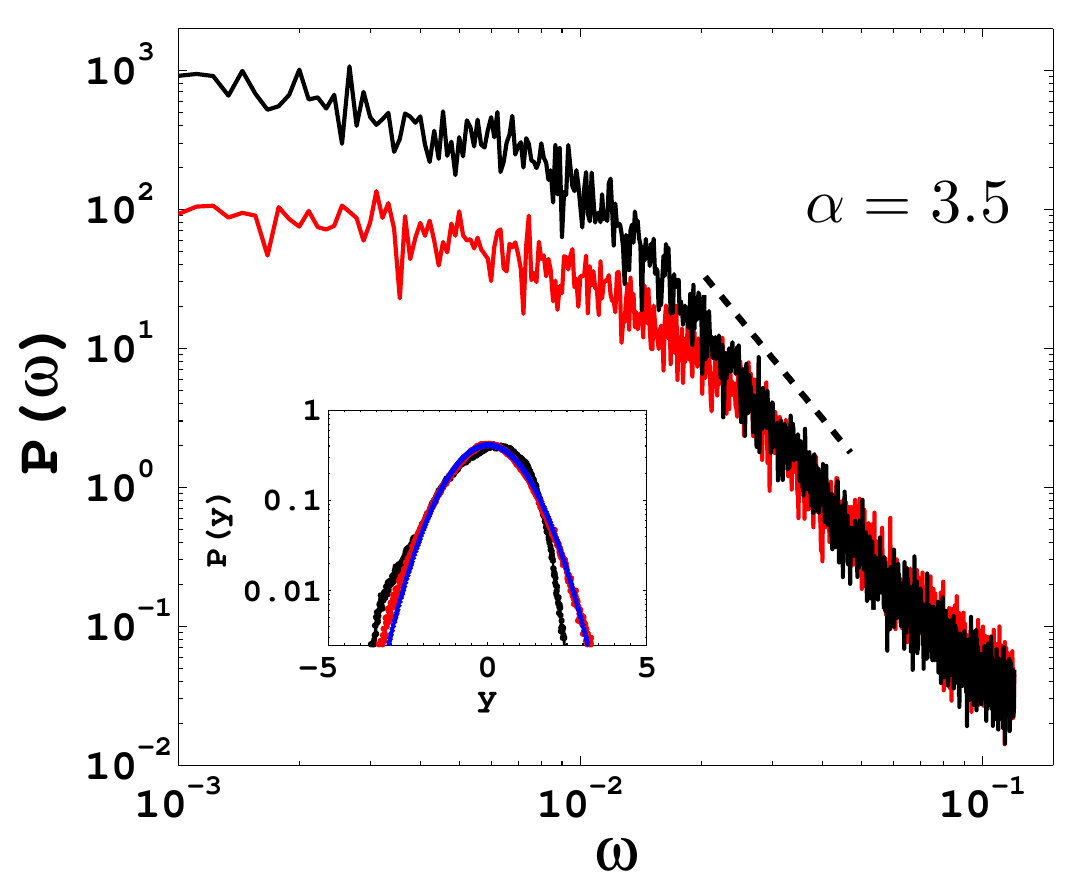}
\caption{Power spectral density in the frequency domain of total injected power ($p(t)$) in shear stress (black) and shear rate (red) controlled simulations for $\lambda_k=1.12$. The dashed line shows a power law decay of the form $P(\omega)\sim \omega^{-\alpha}$ with $\alpha=3.5$. The inset shows the probability distribution function of the normalised fluctuation ($y$) of the total injected power (see text) for both shear rate (red) and shear stress (black) controlled methods and the solid blue line is a gaussian fit. The probability distribution function clearly shows non-gaussianity and negative skewness.}
\label{fig:4}
\end{figure}

\begin{figure}
\includegraphics[width=8.5cm,height=6cm]{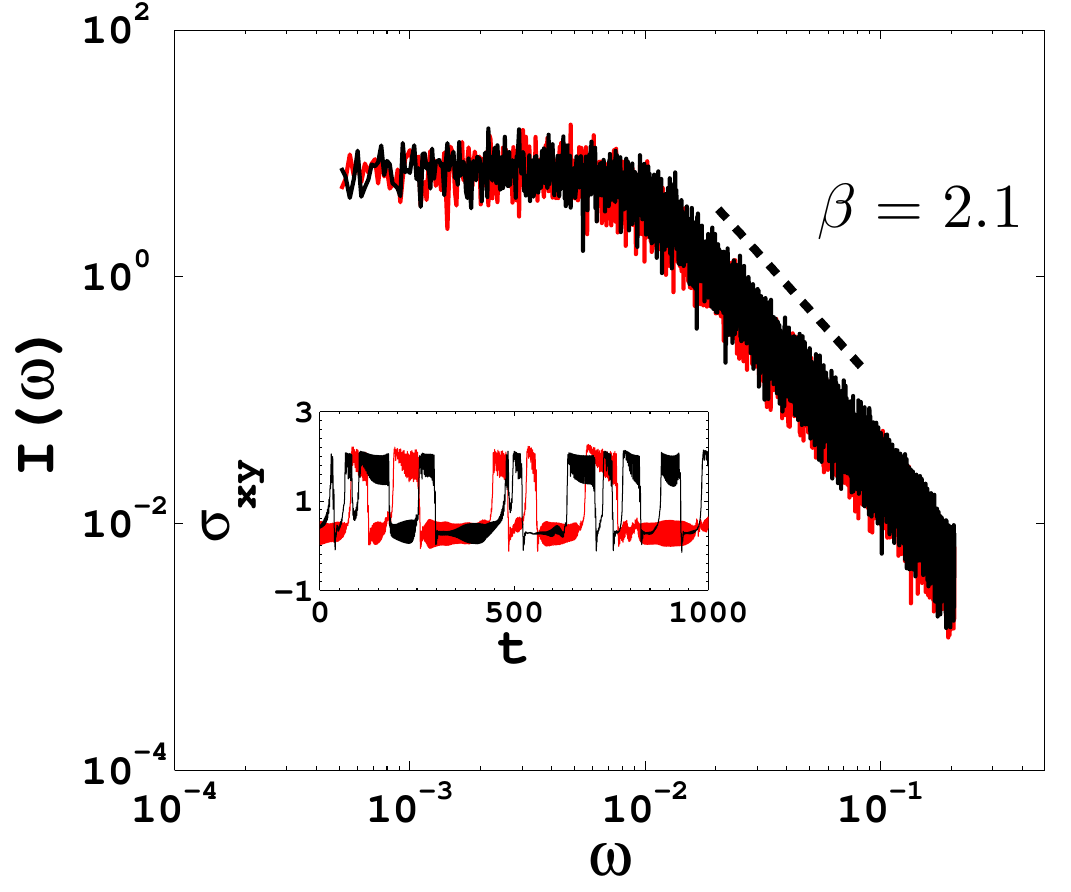}
\caption{Power spectral density in the frequency domain of the order parameter stress, averaged over  $10$ space points (separated by distances greater than the correlation length), in both shear stress (black) and shear rate (red) controlled simulations for $\lambda_k=1.12$. The dashed line shows a power law decay of the form $I(\omega)\sim \omega^{-\beta}$ with $\beta=2.1$. The inset shows the time variation of the order parameter stress in shear rate (red) and shear stress (black) controlled simulations at a fixed space point.}
\label{fig:5}
\end{figure}

We have analyzed the statistical properties of the space and time series of the order parameter stress (Eq.~\ref{eq:OPstress}) and the time series of the total injected power $p(t)= \Gamma {\sigma}_{xy}$. Fig.~\ref{fig:3} shows the power spectrum of the spatial variation of the order parameter stress as a function of the wave number $k$ and a power law scaling behavior $I(k) \sim k^{-\nu}$ with $\nu \approx 2.1$, which is observed for both control protocols in the spatiotemporally chaotic phase. Fig.~\ref{fig:5} shows the power spectrum of the order parameter stress in the frequency domain $\omega$ in the spatiotemporally chaotic phase. Here also a power law scaling behavior is observed, \textit{i.e.} $I(\omega) \sim \omega^{-\beta}$ with $\beta \approx 2.1$ for both control protocols.  Experiments measuring reflected light intensity (which captures the director configuration and thereby is an indirect measure of the order parameter stress) at a given time instant as a function of space or at a fixed space point as a function of time, show scaling behavior with similar exponents~\cite{Fardin:10,Sood:11}. The total injected power $p(t)$ is a fluctuating global quantity whose power spectral density shows a power law decay, $P(\omega) \sim \omega^{-\alpha}$ with the exponent $\alpha \approx 3.5$ for both control protocols as shown in Fig.~\ref{fig:4}. This exponent value is close to the value ($\sim 4$)
found~\cite{Labbe:99,Steinberg3,Steinberg4,Sood:11} for the power spectrum of
the injected power in the frequency domain for elastic turbulence.
The probability distribution of the normalized fluctuation of the total injected power, defined as $y=\frac{p-\langle p\rangle}{\sigma_{p}}$, where $\langle p\rangle$ is the mean value and $\sigma_{p}$ the standard deviation of the instantaneous injected power $p(t)$, shows negative skewness and non-Gaussian behavior (see Fig.~\ref{fig:4} inset). Similar behavior has been observed in experiments on elastic turbulence in polymer solutions~\cite{Labbe:99,Steinberg3,Steinberg4} and micellar solutions~\cite{Sood:11}.

\begin{figure}
\includegraphics[width=8.5cm,height=6cm]{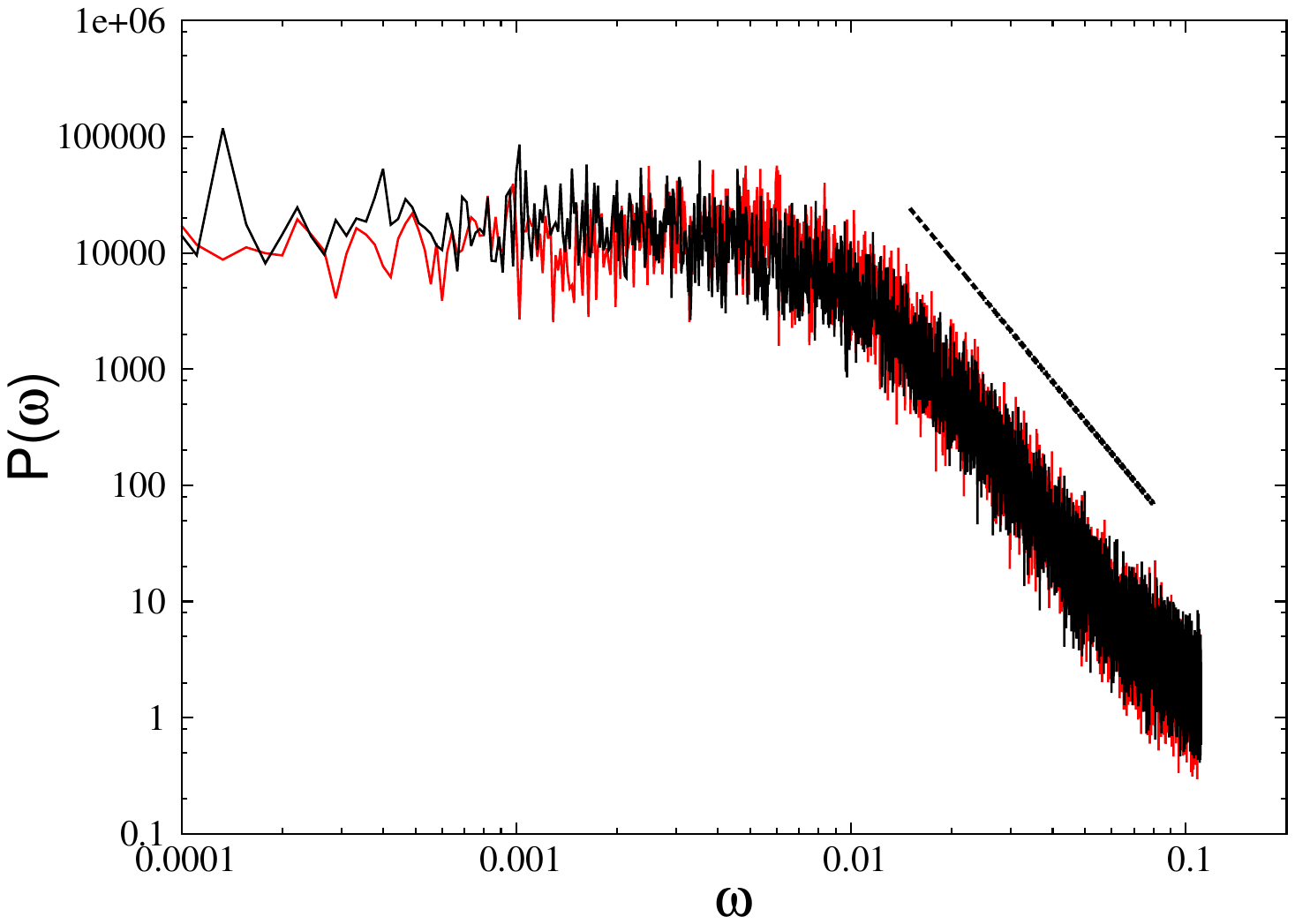}
\caption{Power spectral density of velocity fluctuations in the frequency domain  in both shear stress (black) and shear rate (red) controlled simulations for $\lambda_k=1.12$. The dashed line shows a power law decay of the form $P(\omega)\sim \omega^{-\gamma}$ with $\gamma=3.5$.}
\label{fig:v1}
\end{figure}

\begin{figure}
\includegraphics[width=8.5cm,height=6cm]{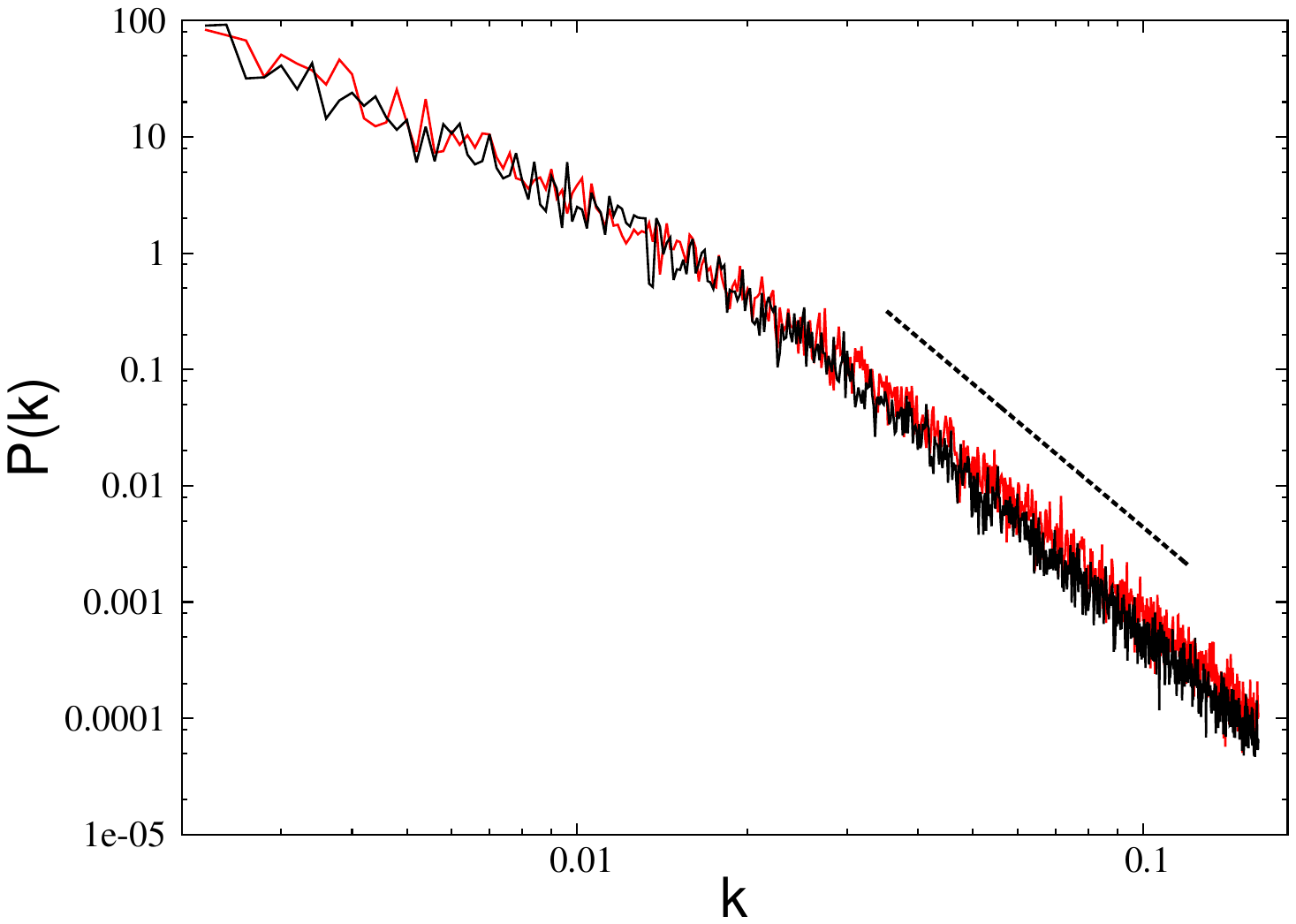}
\caption{Power spectral density of velocity fluctuations in the wavenumber domain in both shear stress (black) and shear rate (red) controlled simulations for $\lambda_k=1.12$. The dashed line shows a power law decay of the form $P(k)\sim k^{-\delta}$ with $\delta=4.1$. }
\label{fig:v2}
\end{figure}

We have also calculated the power spectral density of velocilty fluctuations in both frequency and
wavenumber domains. The results, shown in Figs. \ref{fig:v1} and \ref{fig:v2}, indicate that these
quantities exhibit power-law dependence on their arguments (frequency and wavenumber) with exponents
close to 3.5 and 4.1, respectively. This behavior is similar to that observed in  experiments~\cite{Steinberg1,Steinberg2} and simulations~\cite{simul1,simul2} on elastic turbulence.

\begin{figure}
\includegraphics[width=7.5cm,height=5.4cm]{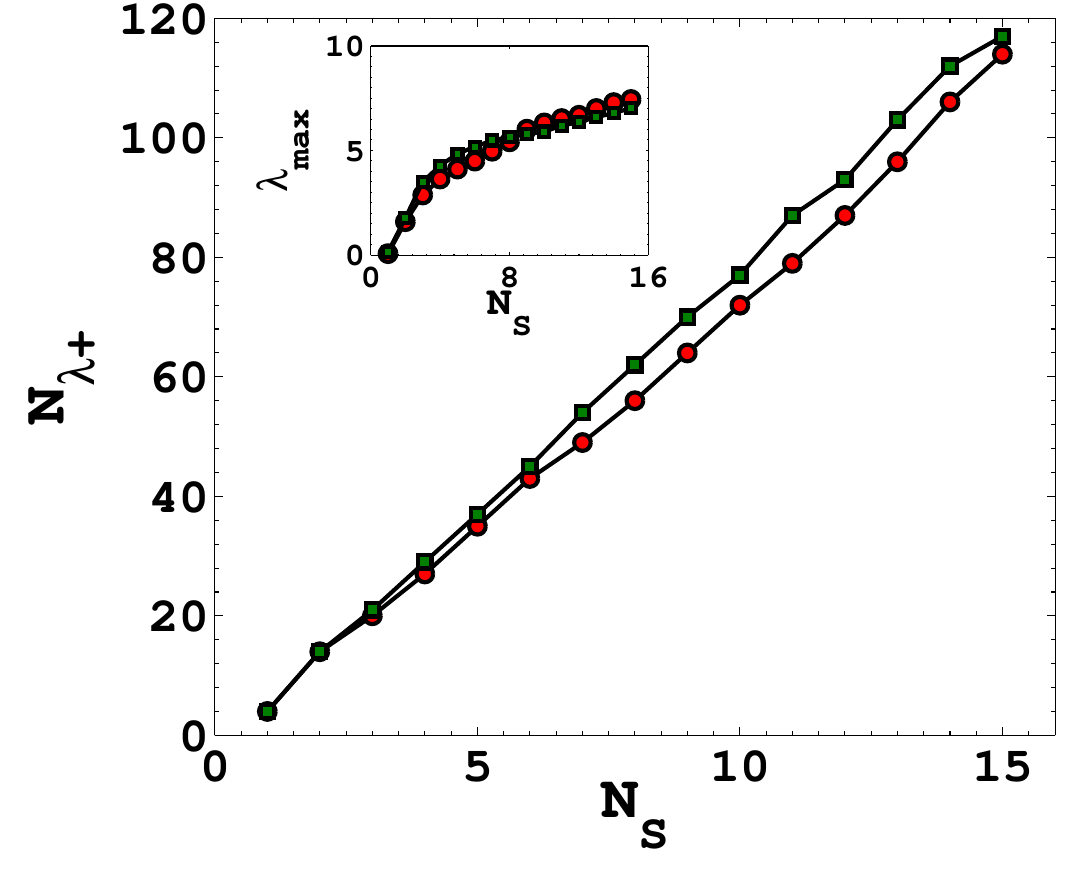}
\caption{Number of positive Lyapunov exponents as a function of sub-system size $N_s$ for shear rate (red) and shear stress (green) controlled methods for $\lambda_k=1.12$. The inset shows the maximum Lyapunov exponent as a function of sub-system size $N_s$ for shear rate (red) and shear stress (green) controlled simulations.}
\label{fig:6}
\end{figure}

To determine the nature of the chaotic state found in our simulations, we have obtained the Lyapunov spectrum~\cite{Schreiber:99} of the order parameter stress time series data. Fig.~\ref{fig:6} shows that the number of positive Lyapunov exponents $N_{\lambda+}$ as well as the maximum Lyapunov exponent $\lambda_{max}$ (see inset) increases as a function of subsystem size $N_{s}$~\cite{Schreiber:99} indicating that the observed fluctuations in the time series data is a signature of underlying spatiotemporally chaotic behavior.

\section{Summary and Discussion}
In conclusion we have numerically analysed the fully non-linear hydrodynamic equations of a sheared nematic fluid under shear stress and strain rate controlled situations incorporating spatial heterogeneity in the gradient direction. We find that the director fluctuations in the unstable region of the flow curve for both controlling protocols show signatures of spatiotemporally chaotic behavior. A detailed analysis of the statistical properties of the fluctuating data train shows resemblance with the behaviour observed in elastic turbulence phenomena in sheared polymer solutions. Though our simple model makes several approximations (\textit{e.g} Stokes limit, incompressibility condition, spatial variation only in the gradient direction \textit{etc.}), the similarity of our results with those of experiments is exciting and it paves the way of extending theoretical work along similar lines to analyse experimental data on driven complex fluids. We hope that our work will spark interest among experimentalists to probe further the connection between rheochaos and elastic turbulence. It would be interesting to carry out numerical investigations on a more realistic model considering spatial heterogeneities in both gradient and vorticity directions. We  expect that such a study will capture the banding instability~\cite{rev1,rev2,rev3} seen in experiments. 

We thank A. K. Sood and Ananyo Maitra for very useful discussions. R. M. acknowledges financial support from CSIR, India. C. D. acknowledges financial support from DST, India. B. C. acknowledges support from Durham University and IISc Bangalore for hospitality.

\end{document}